\documentclass[10pt, a4paper,twocolumn]{article}

\usepackage{graphicx}
\usepackage{amsmath}
\usepackage{amsfonts}
\usepackage{amssymb}
\usepackage{hyperref}

\begin{document}

\twocolumn[
  \begin{@twocolumnfalse}
\def\lsim{\mathrel{\raise.3ex\hbox{$<$\kern-.75em\lower1ex\hbox{$\sim$}}}}
\def\gsim{\mathrel{\raise.3ex\hbox{$>$\kern-.75em\lower1ex\hbox{$\sim$}}}}
\sf
\centerline{\Huge Proton decay and grand unification$^{\S}$  }
\vspace{5mm}
\centerline{\large Goran Senjanovi\' c}
%\vspace{5mm}
\centerline{{\it International Centre for Theoretical Physics, 34100 Trieste, Italy }}
\vspace{5mm}
\centerline{\large\sc Abstract}
\begin{quote}
\small
  I review the theoretical and experimental status of proton decay theory and experiment.
  Regarding theory, I focus mostly, but not only, on grand unification. I discuss only the minimal,
  well established SU(5) and SO(10) models, both ordinary and supersymmetric. I show how the
  minimal realistic extensions of the original Georgi - Glashow model can lead to interesting
  LHC physics, and I demonstrate that the minimal supersymmetric SU(5) theory is in perfect
  accord with experiment.  Since no universally accepted model has of yet emerged, I discuss
  the effective operator analysis of proton decay and some related predictions from a high scale
  underlying theory. A strong case is made for the improvement of experimental limits, or better 
  the search of, two body neutron decay modes into charged kaons and charged leptons. Their
  discovery would necessarily  imply a low energy physics since they practically vanish in any
  theory with a desert in energies between $M_W$ and $M_{GUT}$.

\end{quote}

 \end{@twocolumnfalse}]
 {
 \renewcommand{\thefootnote}%
   {\fnsymbol{footnote}}
 \footnotetext[4]{Based on the plenary talks at the SUSY09 and  PASCOS09 conferences.}
}

 \rm
\newpage

\section{Introduction}

  As Maurice Goldhaber  had put nicely more than half a century ago,  we feel it in our bones that proton 
  is long lived, for otherwise the radiation from the decay would kill you. All you need to know is how much
  radiation is hazardous for you, and you get a lower limit on proton lifetime, on the order of
  $10^{18}$ years.  It was, as it is today, puzzling  that the proton should live so much longer
  than the neutron, and it was attributed to the conservation of the baryon number, first by Weyl \cite{Weyl:1929fm}
already in 1929, and the again by Stuckelberg \cite{stuek} and Wigner \cite{wigner}every ten years afterwards.  
And then it was becoming a dogma, as usually happens when a cause and a consequence are confused, and the questioning
became heretic. It is amusing to see how almost apologetic Reines et al \cite{Reines:1954pg} were in 
justifying their pioneering experimental probe of nucleon decay:

"It has often been surmised that there exists a conservation law 
of nucleons, i.e., that they neither decay spontaneously nor are destroyed or created singly in nuclear collisions.
In view of the fundamental nature of such an assumption, it seemed of interest to 
investigate the extent to which the stability of nucleons could be 
experimentally demonstrated." 

They established a limit of $10^{21}$ yr, which they would improve to $10^{26}$ yr later on. 
And while the baryon number was becoming sacred, lepton number violation was seriously 
discussed already in 1937 by Majorana in his classic paper \cite{Majorana:1937vz} on majorana
spinors. In turn Racah  \cite{Racah:1937qq} and Furry  \cite{Furry:1939qr} were to discuss at depth neutrino-less double decay still desperately
searched for as a probe of lepton number violation. This shows how crucial for experimentalists it is to have 
a theory behind.

  Since there is nothing sacred about global symmetries 
we will follow a belief that:  {\bf the only good global symmetry is a broken global symmetry}. 
 The symmetry point is not a special point in the parameter space of a theory, 
 for an almost exact global symmetry is just as protected and useful as a fully
exact one. Furthermore a zero is useless for experimentalists. 

   And baryon number is not an exact gauge symmetry unless you accept a corresponding gauge coupling 
   to be ridiculously  small:  $g_B \leq (10^{-20} - 10^{-19})$  (a repulsive competition with gravity). 
  
    While Goldhaber et al had to justify their search for proton decay, today the atmosphere has changed completely,
    as illustrated by the fact that one is asked to give plenary talks on it at major conferences. The point is 
      that there is a theory behind: 
  grand unification of strong, weak and electro-magnetic interactions.  It grew out of pioneering ideas of 
  Pati and Salam \cite{patisalam} on the unification of quarks and leptons and is exemplified perfectly on the minimal SU(5) 
  theory of Georgi and Glashow \cite{Georgi:1974sy}.   When Georgi, Quinn and Weinberg  \cite{Georgi:1974yf}
  computed the unification scale
  and predicted proton lifetime $\tau_p \simeq 10^{30}$ yr, many experimentalists rushed underground 
   \cite{Goldhaber:1988gb},  
  all over the world, from India to Japan to US to Europe.  Here is a list of experiments
  
  {\bf Calorimeter detector}
\begin{itemize}
\item
Kolar Gold Field - Kolar district (Kamataka, India) 

\item
NUSEX - Mont Blanc (Alps, France) 

\item
FREJUS - Frejus tunnel (Alps, France) 

\item
SOUDAN- Soudan underground mine (Minnesota, US) 
\end{itemize}

  {\bf Cherenkov detector}
\begin{itemize} 
\item
IMB - Morton salt mine (Ohio, US) 

\item
Kamiokande - Mozumi mine (Hida, Japan)  $\to$ Super-Kamiokande
$\to$ atmospheric neutrino oscillations
\end{itemize}

  The search resulted in great improvements on proton decay limits, culminating with SK 
that recently pushed the pionic channel all the way to about $10^{34}$ yr. Here are limits
on some proton decay channels.

\vskip 0.5cm
\begin{tabular}{lc}
  Channel &    $\tau_p(10^{33}$ years) \\
 & \\
$p \to e^+  \pi^0 $ &                                                       8.2 \\

$p \to  \mu^+  \pi^0 $   &                                                     6.6\\

$p \to \mu^+ K^0 $  &                                                            1.3 \\

$p \to e^+ K^0 $      &                                                        1.0\\
 
$p \to  \nu K^+ $       &                                                               2.3 \\
 
$n \to e^+ K^- $      &                                                        0.02\\
 
$n \to e^- K^+$      &                                                        0.03\\
  
\end{tabular}
\vskip 0.5cm
 
The last two channels  are important since they are an indication of low  scale as discussed in sections
3 and 5. 

  In what follows I discuss the minimal SU(5) and SO(10) theories, both ordinary and supersymmetric. Due
  to the lack of space, many important references are likely to be omitted. For a review and more complete
  references on the subject, see \cite{Nath:2006ut}.

\section{ \hspace{-0.7cm}  \Large{Minimal realistic SU(5)}   }

  It was the minimal SU(5) that caused the underground rush, for it predicted proton lifetime on the order
  of $\tau_p \simeq 10^{30}$ yr.  And on top, it also predicted the nucleon decay branching ratios as shown
  by Mohapatra \cite{Mohapatra:1979yj}; unfortunately these predictions resulted from the wrong mass relations:
  $m_e = m_d$, wrong for all three generations. These relations can be corrected easily by simply adding higher
  dimensional operators  \cite{Ellis:1979fg} (at least for the first two generations) , but then the theory stops being predictive. 
 
   In any case this is only history now, for the theory is not even consistent:

\begin{itemize}
\item
gauge couplings do not unify since $\alpha_2$  and $\alpha_3$ meet at $10^{16} \mbox{ GeV}$ 
(as in SM), but $\alpha_1$ meets $\alpha_2$ too early at $\approx 10^{13}  \mbox{ GeV}$ ;

\item
neutrinos are massless as in the SM.

\end{itemize}

   Possible higher dimensional operators are not enough: neutrino mass comes out too small ($\lesssim 10^{-4} eV$)
   and the threshold effects do not cure the lack of unification.

    It is important to know then what the minimal consistent realistic extensions are.
   There are two. You can
   
   \begin{itemize}
   
   \item add a symmetric complex scalar field (and higher dimensional operators for charged fermion masses)
   \cite{Dorsner:2005fq}
   $15_H = (1_C, 3_W) + (6_C, 1_W) + [(3_C, 2_W) = leptoquarks] $, with $(1_C, 3_W)$ being the usual 
     Higgs triplet behind the type II seesaw  \cite{Magg:1980ut}.
    The leptoquarks $(3_C, 2_W)$ may remain light (but not necessarily), and a rather interesting prediction
       is a fast proton decay, on the edge of experimental limits.
       
       \item add an adjoint fermion field
         $24_F=  (8_C, 1_W) + (1_C, 3_W) + (1_C, 1_W) + (3_C, 2_W) + (\bar 3, 2_W)$.
        The fields   $(1_C, 1_W) + (1_C, 3_W)$ are responsible for type I \cite{seesaw} + III \cite{Foot:1988aq} hybrid seesaw. The model requires
       higher dimensional operators both for charged femions and for realistic neutrino Dirac Yukawa couplings
It predicts a light fermion triplet  $(1_C, 3_W)$, with a mass below TeV so that the running of the SU(2) 
 gauge coupling is slowed down and meets U(1) above $10^{15} \mbox{ GeV}$.  It's phenomenology is quite interesting for
 it leads to lepton number violation at colliders in the form of same sign di - leptons as suggested originally in seesaw
 a long time ago \cite{Keung:1983uu}. For the relevant studies in the context of the type III seesaw 
 see \cite{Franceschini:2008pz}.
 % \cite{Arhrib:2009mz}.

\end{itemize}

   The two theories have in common a `fast' proton decay, with $\tau_p \leq 10^{35}$ yr, to keep in mind in what
   follows.

\section{\hspace{-0.5cm}   \Large{Minimal supersymmetric SU(5)}    }

 The underground rush continued, or better to say, got boosted with the success of the minimal supersymmetric 
 SU(5).  Low energy supersymmetry, suggested in order to stabilize the Higgs mass hierarchy, predicted correctly
 $sin^2 \theta_W = 0.23$ in 1981 \cite{Dimopoulos:1981yj} \cite{Ibanez:1981yh} \cite{Einhorn:1981sx}
 \cite{Marciano:1981un}
 ten years before its confirmation at LEP. It actually did even better:  the prediction
 of $sin^2 \theta_W = 0.23$ was tied to the prediction of the heavy top quark, with $m_t \simeq 200  \mbox{ GeV}$.  Namely,
 in 1981 the low indirect measurements gave $sin^2 \theta_W = 0.21$, with the assumed value $\rho = 1$.  In order
 to make a case for low energy supersymmetry, Marciano and I \cite{Marciano:1981un}
 had to say that  $\rho$ was bigger, which required
 loops, which required at least one large coupling, and a natural SM candidate was the top quark, with $y_t \simeq 1$.
 It is remarkable that both the $sin^2 \theta_W = 0.23$ and the heavy top would turn out to be true.  It should be stressed
 that heavy top quark played also a crucial role in the radiative Higgs mechanism \cite{Inoue:1982pi} \cite{Alvarez-Gaume:1983gj}.  Thus heavy top is an integral part of low energy supersymmetry.
 
    The GUT scale was predicted:  $M_{GUT} \simeq 10^{16} \mbox{ GeV}$ and in turn {$\tau_p (d=6) \simeq 10^{35 \pm 1}$ yr,
    which would have rendered proton decay out of experimental reach.
However, supersymmetry leads to a new contribution:  $d=5$ operators \cite{Sakai:1981pk}
through the exchange of heavy color triplet Higgsino ($T$ and $\bar T$). A rough estimate gives

   $$ G_T   \simeq \frac{\alpha}{4 \pi} y_u \, y_d \frac{m_{\mbox{\tiny gaugino}}}{M_T m^2_{\tilde f}} \simeq 10^{-30} \mbox{ GeV}^{-2} $$
 which for 
$y_u \simeq y_d \simeq 10^{-4}$, $m_{\mbox{\tiny gaugino}} \simeq 100 \mbox{ GeV} $, 
$m_{\tilde f} \simeq $TeV and $M_T \simeq 10^{16} $ GeV gives
$\tau_p (d=5) \simeq 10^{30 - 31}$ yr.  It would seem that today this theory is ruled out. It was actually proclaimed
dead in 2001  when the triplet mass was carefully computed to give $M_T^0 = 3 \times 10^{15} \mbox{GeV}$
\cite{Murayama:2001ur} (for the superscript 0 explanation, see below).
 Caution must be raised however for two important reasons:
i) the uncertainty in sfermion masses and mixings \cite{Bajc:2002bv} and ii) uncertainty in $M_T$ \cite{Bajc:2002pg}
due to necessity
of higher dimensional operators \cite{Ellis:1979fg} to correct bad fermion mass relations $m_d = m_\ell$ 
\cite{Chanowitz:1977ye}.
The $d=4$ operators, besides correcting these relations also split the masses $m_3$ and $m_8$ of weak 
triplet and color octet, respectively, in the adjoint $24_H$ Higgs super multiplet and one gets

$$
\left.
\begin{array}{l}
M_{GUT} = M^0_{GUT} \left(  \frac{M_{GUT}^0}{ 2 m_8}\right)^{1/2}  \\ M_T = M_T^0 \left(\frac{m_3}{m_8} \right)^{5/2}
\end{array} \right\}
\begin{array}{r}
M^0_{GUT} \simeq 10^{16} \mbox{ GeV}\vspace{3mm} \\
M_T^0 = 3\times 10^{15} \mbox{ GeV}
\end{array}$$

where the superscript 0 denotes the predictions for $m_3=m_8$ at the tree level with $d=5$ operators
neglected. The fact that $M_{GUT}$ goes up with $m_8$ below $M_{GUT}$ was noticed quite some time ago
\cite{Bachas:1995yt}.
Imagine that
 $d=4$  terms dominate for small cubic Yukawa self coupling, in which case one has $m_3 = 4 m_8$ and thus
$M_T = 32 M_T^0 \simeq 10^{17} \mbox{ GeV} \simeq M_{GUT} $ 
$(m_8 \simeq 10^{15} \mbox{ GeV})$. In turn a strong suppression of proton decay with 
$ \tau_p \simeq 10^3 \tau_p^0 (d=5) \simeq 10^{33-34}$ yr.  In principle the ratio of the triplet and octet masses
can be as large as one wishes, so at first glance the proton lifetime would seem not to be limited from above at all.
However, all this makes sense if the theory remains perturbative and thus predictive. Increasing $M_{GUT}$ would 
bring it too close to the Planck scale, so it is fair to conclude that the proton lifetime is below $10^{35}$ yr. This 
prediction is not hard, though. The d=4 operators not only cure the bad mass relations, but also also split the
Yukawa couplings of the SM doublet Higgs and the color triplet. if you allow for cancellation between the d=3
and d=4 couplings, one can in principle make the color triplet couplings as small as one wishes and thus 
suppress the proton decay. In principle, it is even possible to make a color triplet mass at the electro-weak 
scale \cite{Dvali:1992hc}.  The cancellation of matrices is rather unnatural, so I will not pursue it here.
 However, it may
emerge in more complex models, such as SO(10) \cite{Dvali:1995hp}, so this should be kept mind as a 
serious possibility.

  In short, the minimal supersymmetric SU(5) is still a perfectly viable theory, and the $d=5$ proton decay is 
  expected close
  to the present limit. The theory is crying for a new generation of proton decay experiments.  It would seem that B-L remains
  an accidental global symmetry of nucleon decay and one expects the dominant mode $p \to K^+ \bar \nu_\mu$
  characteristic of $d=5$ operators. However, this minimal theory must account for neutrino masses and mixings which
  implies that R-parity is broken. 
     The first important implication of not assuming R-parity  is that the lightest neutralino cannot be dark matter,  for it
     decays too fast with the collider signature of lepton number violation. Thus, the only dark matter candidate is 
     an unstable gravitino.   It decays into the neutrino and the photon through the neutrino-gaugino mixing
     
$$ \Theta_{\nu \, \mbox{\tiny gaugino}} \simeq \sqrt{\frac{m_\nu}{M_{\mbox{\tiny gaugino}}}}
$$

   The decay is suppressed by the Planck scale \cite{Takayama:2000uz}
\begin{eqnarray}
\Gamma (3/2 \to \gamma \nu) &=& \frac{1}{32 \pi} \frac{m_{3/2}^3}{M_{Pl}^2} \Theta^2_{\nu \, \mbox{\tiny gaugino}} \nonumber\\& & \simeq  \frac{1}{32 \pi} \frac{m_{3/2}^3}{M_{Pl}^2} \frac{m_\nu}{M_{\mbox{\tiny gaugino}}} \nonumber \\
& &  \leq  10^{-50} \mbox{ GeV}  \;\;\;\;\; (\tau \geq 10^{26} sec). \nonumber
\end{eqnarray}

 The lower limit on gravitino lifetime comes from requiring that the flux of produced photons does not exceed the observed  diffuse photon background, and it gives an upper limit on gravitino mass, 
 $ m_{3/2} \leq 1-10 \mbox{ GeV}$ \cite{Takayama:2000uz}, reasonable for the LSP. If one relaxes the neutrino-gaugino
 mixing, gravitino can be heavier; see e.g. \cite{Ishiwata:2008cu}. For a review and references on gravitino dark matter,
 see \cite {Buchmuller:2009fm}.   
%$\Rightarrow$ \shadowbox{$ m_{3/2} \leq 10 \mbox{ GeV$ }

      Once  R-parity is not assumed ad hoc, one has a new source of proton decay too. Due to allowed couplings
      in the superpotential
 $ \lambda_1 \, u^c \, d^c \, d^c + \lambda_2 \, q \, \ell \, d^c  + \lambda_3 \, \ell \, \ell \, e^c $,
$\tilde{d^c} $ mediates $d=6$ proton decay, which implies the limit
   $\lambda_1 \, \lambda_2 \leq 10^{-25}  \; (= ?)$   for $m_{\tilde{d^c}} \simeq TeV$. The question mark indicates
   an interesting possibility of these couplings causing proton decay. It can be shown that the parameter space allows
   for a B+L violating mode $n \to e + K^+$ \cite{Vissani:1995hp}. As I show below, this decay mode cannot come from
   a conventional picture of grand unification with a desert.

\section{SO(10)}

       Although SU(5) is the minimal theory of grand unification and as such deserves maximal attention as
       a laboratory for studying  proton decay, SO(10) has important merits

\begin{itemize}

 \item it unifies a fermion family in a spinorial $16_F$ representation and as such is a minimal unified theory 
of matter and interactions
   \item it automatically contains right-handed neutrinos   $N$
 
   \item it gives naturally  $ M_N \gg M_W$ and so neutrino has a tiny mass through the see-saw mechanism

 \item in supersymmetry R-parity is a gauge symmetry \cite{rparity}

%Matter parity \az{$\Leftrightarrow$} \az{$R(p)$ $\in$ center of $SO(10)$}

%\cita{Mohapatra '86}

 \item in the renormalizable version R-parity
 remains exact \cite{Aulakh:1997ba}
 and the lightest supersymmetric partner (LSP ) is stable, and thus becomes a natural
 dark matter candidate.

%\cita{Aulakh, Benakli, G.S. '96 ; Aulakh {\em et al.} '00}

\end{itemize}

While ordinary, non supersymmetric SO(10) was studied at length over the years, no predictive realistic model
of fermion masses and mixings ever emerged.  

All the fermion
masses and mixings can be accounted for with the $10_H$ and $126_H$ representations, the latter providing
a large mass to the right-handed neutrinos. In the Pati-
Salam $SU(2)_L \times SU(2)_R \times SU(4)_C$ language 
$$10_H = (2, 2, 1) + (1,1, 6)$$  
and 
$$126_H= (2, 2, 15) + (3, 1,10) + (1, 3, \bar 10) + (1, 1, 6 )$$
and one can see that in principle no bad mass relations come out.  If $10_H$ is taken to be real, a minimal scenario,
the predictions turn out to be wrong \cite{Bajc:2005zf}, and making it complex kills the predictions.  This is why 
ordinary SO(10) does not do the job. The theory deserves attention for it allows for an intermediate L-R symmetry
\cite{leftright}; for recent attempts to revive ordinary SO(10) see \cite{Bajc:2005zf}, \cite{Bertolini:2009qj}.

This situation improves in supersymmetry: the  couplings being holomorphic simplifies things.  Although a single
$10_H$ is necessarily complex, still, analyticity guarantees a single Yukawa, and similarly for $126_H$. This
minimal supersymmetric version,  coined renormalizable,  although suggested already in 1982 \cite{susyso10},
and revisited ten years later \cite{Babu:1992ia}, has been studied at length only in recent years \cite{japon}
 \cite{moresusyso10}.
A boost was provided by an observation that $b - \tau$ unification can be naturally
tied with the large atmospheric mixing angled in the type II seesaw \cite{Bajc:2002iw}.  This means that
small quark and large lepton weak mixing angles follow naturally from the common Yukawa couplings without
any need for flavor symmetries. This is an important result which shows that the much talked about issue
of rather different quark and lepton mixings is not a problem as normally argued. In the SO(10) grand unified
theory it is simply a product of a broken quark-lepton or Pati-Salam symmetry. In other words, although 
at the GUT scale quarks and leptons are completely equivalent with the same Yukawa couplings, at the
SM energies, their different masses lead to different mixing angles.

After a great initial success,  when pinned down, the theory ran into tension between fast proton 
and neutrino masses. It was revisited recently \cite{Bajc:2008dc}
and shown to work with the so called split supersymmetry
spectrum of heavy sfermions. It is important to note that the nucleon decay branching ratios are determined.
This is an example for a kind of theory of proton decay we are searching for.
It remains to be shown that the solution found is unique.

   Instead of a large $126_H$ Higgs, one may choose a $16_H$ and then build $126_H$ Higgs effectively as        
$16_H^2$. This induces a proliferation of couplings and one must introduce extra flavor symmetries, and thus go
beyond a simple GUT picture.  For a discussion of this approach, see \cite{babu}. In any 
case, it is clear that no accepted minimal model has yet emerged. In this sense it is really important to keep in mind
that the minimal supersymmetric SU(5) theory is not ruled out. It should be viewed as the laboratory for studying grand 
unification and related issues, such as proton decay and magnetic monopoles, the way the minimal ordinary
SU(5) used to be when it worked.

\section{Effective operator analysis of nucleon decay and the desert picture}

 Since we do not have {\bf the} theory of grand unification it is worthwhile to study the generic features of 
 high scale theories of proton decay. This is done through the effective operator expansion \cite{Weinberg:1979sa}.
 In this program one
 expands the baryon violating operators in $M_W/M_B$ or $m_p/M_B$, where $M_B$ is the scale responsible for proton decay.
 If $M_B$ is very large,   one can safely assume the SM
 gauge symmetry unbroken at that scale. This is surely true  in
 conventional grand unification with a desert where $M_B = M_{GUT}$.
     
     There are only four leading $d=6$ operators \cite{Abbott:1980zj}
\begin{center}
\hspace{0cm}  $O_1=$ $(u_R \, d_R)\:(q_L \, \ell_L) $   \hskip  1.5 cm   $O_2=$  $(q_L \, q_L ) \; (u_R \, e_R) $\\
 \hspace{0cm}  $O_3=$  $(q_L \, q_L ) \; (q_L \, \ell_L) $    \hskip 1.5 cm  $O_4=$ $(u_R \, d _R   ) \; (u_R \, e_R)$
\end{center}

  The gauge meson dominance would imply only the first two which can lead to a number of predictions. 
Even in general one has an immediate prediction of an accidental B - L symmetry  in any theory with a high
scale $M_B$, which explains why any GUT model was predicting it.

 There are also a number of immediate isospin relations
  \begin{center}
  $\Gamma(p\to \ell_R^+ \, \pi^0) = \frac{1}{2} \Gamma (n \to \ell_R^+ \, \pi^-) = \frac{1}{2} \Gamma(p\to \bar \nu \pi^+) = \Gamma(n \to \bar\nu \, \pi^0)$
\end{center}
\hskip 1.9 cm {$\Gamma(p\to \ell_L^+ \, \pi^0) = \frac{1}{2} \Gamma(n \to \ell_L^+ \pi^-)$

Also, it is evident that  $\bar s$ goes out, such as in the allowed decay $p\to K^+ \bar \nu$ , characteristic of d=5 operators
in supersymmetry.  

 One conclude in turns that there can be no two body neutron decay into kaons and charged leptons
   \begin{center}
% \shadowbox{ 
 $n \not\to K^+ \, \ell$
  % \shadowbox{ 
  \hskip 1.5 cm
  $n \not\to K^- \, \ell^+$
\end{center}

  This is why the R-parity violating mode $n \to K^+ \, e$ discussed in section 3. is so important: if discovered,  it
  would point out immediately towards a low scale source of proton decay. It results from the operator
  
  $$d\, s\, d\, \bar\ell \, \langle H \rangle / \tilde m^3$$
  
  where the Higgs vev is needed to break the SM symmetry which
  otherwise guarantees the conservation of B - L.  The low scale $\tilde m$ of supersymmetry breaking allows for
  this operator not being suppressed.
  In GUT $\tilde m$ becomes $M_{GUT}$, implying the  suppression $M_W/M_{GUT}$. Furthermore, the necessary presence of s
  quark for symmetry reasons implies the absence of the pionic mode:  $n \not\to \pi^+ \, e$, which makes the it
  even more predictive \cite{Vissani:1995hp}.
  
    This leads to an important message: the decay modes
    $n \to K^+ \, e $   and $n \to K^- \, e^+ $ 
 imply a low energy source of proton decay. The limits are roughly $10^{31}\mbox{ yr}$ and I 
    urge the experimentalists to improve them.

\section{Proton decay:  matrix elements}

  Due to the shortage of space, I will be very brief here.  There have been many attempts of
  computing the matrix elements using the non-relativistic quark model, the bag model, the lattice and the
  chiral Lagrangian  techniques.  In recent years, the focus has been on the lattice and chiral Langrangians, 
especially the combination of the two.  For the work on lattice  see e.g. \cite{Aoki:2006ib}.
A lot of progress was made on the lattice computation of chiral Lagrangian coefficients, see e.g.  \cite{Aoki:2008ku}.
The trouble is that chiral perturbation theory works great for soft pions, while the pions  from the proton decay
would have the momentum on the order of proton mass. Thus, one needs  to go beyond the leading term
of the original classic \cite{Claudson:1981gh} .

\section{Summary and Outlook}
  
     I would conclude with the following messages:
     
     \begin{itemize}
     \item there is no universally accepted grand unified theory of nucleon decay

%%\vskip 0.2cm
\item most models give$\tau_p \leq 10^{35}$ yr,  especially with low energy supersymmetry
%%\vskip 0.2 cm
\item
minimal SU(5) is ruled out and its minimal extension leads to possible LHC physics
through the light fermion triplet
%\vskip 0.2cm

\item

minimal supersymmetric  SU(5) still perfectly viable with gravitino being the only possible (unstable) dark matter
%\vskip 0.2 cm

\item test of GUT desert picture provided by decay modes
 $n \to K^+ \, \ell$  and $n \to K^- \, \ell^+$ 
which would indicate low energy physics as a source of p decay,
such as R-parity violating couplings

%$\Rightarrow$ connection between p decay and LHC

\end{itemize}

 Thus,  a new generation of experiments is badly needed.
 Here is a list of proposals
\vskip 0.3cm

{ \bf Cherenkov  detector} - MEGATON (hopefully):  good for the pion modes
%\vskip 0.2 cm

\begin{itemize}

\item
HYPER-Kamiokande 

%\vskip 0.2cm

\item
3M -(Megaton, Modular, Multipurpose)

 Homestake - DUSEL (Deep Underground Science and Engineering Lab)

%\vskip 0.2 cm
\item
MEMPHYS - MEgaton Mass PHYSics

Fr\'ejus - LAGUNA (Large Apparatus Grand Unification and Neutrino Astrophysics)  project

%\vskip 0.2cm

%\begin{center}
 % good for the $p \to \ell^+ \pi^0$ mode
 % \end{center}

\end{itemize}

{\bf Liquid Argon Detector} - 100 kT:   good for the kaon modes
%\vskip 0.2cm

\begin{itemize}

\item
LANNDD  (Liquid Argon Neutrino Nucleon Decay Detector)

 Homestake - DUSEL?

\item

GLACIER  (Giant Liquid Argon Charge Imaging ExpeRiment)

Europe - Laguna

\end{itemize}
%\vskip 0.2cm

{\bf Liquid scintillator} - 50 kT:  good for the kaon modes
\begin{itemize}
\item
%\vskip 0.2cm
LENA (Low Energy Neutrino Astronomy)

 Europe - Laguna

\end{itemize}

%%\vskip 0.2 cm

%%\vskip 0.3cm
 
Hopefully they all will be funded and some will reach $10^{35}\mbox{ yr}$ in 10-20 years? One cannot overemphasize the
importance of trying to reach this scale, generic of grand unification. Without experiment, a beautiful field of proton decay
and grand unification is bound to turn into metaphysics.

%%%%%%%%%%%%%%%%%%%%%%%%%%%%%%%%%%%%%%%%%%%%%%%%
%% BACKMATTER
%%%%%%%%%%%%%%%%%%%%%%%%%%%%%%%%%%%%%%%%%%%%%%%%

\section{Acknowledgements}

  I wish to thank Pran Nath  and other organizers of SUSY 09 for an excellent conference. I also
  wish to thank Wilfried Buchm\"uller and Laura Covi and others at DESY for another great
  meeting, PASCOS 09. This talk was ordered by the organizers of both as a review of proton decay and
  grand unification.  I am
  deeply grateful to my collaborators on the topics discussed here Charan Aulakh, Borut Bajc,
  Pavel Fileviez P\'erez, Bill Marciano, Alejandra Melfo, Rabi Mohapatra, Miha Nemev\v{s}ek,  Andrija Ra\v{s}in (Andrija, vrati se, sve 
  ti je opro\v{s}teno) and Francesco Vissani. 
 
 I  am grateful to Ilja Dor\v{s}ner, Gia Dvali, Alejandra Melfo and Enkhbat Tsedenbaljir for useful discussions. 
  Thanks are also due to Vladimir Tello for his help with this manuscript. I am deeply grateful to Borut Bajc, Svjetlana
  Fajfer and other members of the theory group at "Jo\v{z}ef  Stefan" Institute in Ljubljana for their warm hospitality during
  this write up.
   This work was partially supported by the EU FP6 Marie Curie Research and Training Network "UniverseNet" (MRTN-CT-2006-035863) and by the Senior Scientist Award from the Ministry of Science and Technology of Slovenia.

%\section*{References}

\end{document}